\documentclass{IEEEtran}

\usepackage[T1]{fontenc} 
\usepackage{microtype}   

\usepackage[keeplastbox]{flushend} 

\usepackage{float}

\usepackage{tikz, tikz-3dplot}
\usetikzlibrary{calc,matrix,positioning,arrows.meta,cd,shapes,shapes.multipart}

\newcommand*\state[1]{\tikz[baseline=(char.base)]{
            \node[shape=rounded rectangle,draw,inner sep=2pt,scale=0.9] (char) {$#1$};}}

\newcommand*\sstate[2]{\tikz[baseline=(char.base)]{
            \node[shape=rectangle split,rectangle split parts=#1, rectangle split horizontal, rounded corners, draw,inner sep=2pt,scale=0.9] (char) {#2};}}

\newcommand*\twostate[2]{\sstate{2}{$#1$ \nodepart{two} $#2$}}

\usepackage{makecell}
\usepackage{multirow}

\usepackage{mathbbol}
\usepackage{amsfonts,amssymb,amsmath,mathtools}
\usepackage{stmaryrd} 
\DeclareSymbolFontAlphabet{\mathbb}{bbold}
\DeclareSymbolFontAlphabet{\mathbbg}{bbold}

\newcommand{\set}[1]{\ensuremath{\mathbf{#1}}}
\newcommand{\powerset}[1]{\ensuremath{\mathcal{P}(#1)}}

\newcommand{\fcn}[1]{\ensuremath{\mathsf{#1}}}

\usepackage{xstring}
\newcommand{\StrHead}[1]{\StrLeft{#1}{1}}
\newcommand{\StrTail}[1]{\StrGobbleLeft{#1}{1}}

\newcommand{\lattice}[1]{\ensuremath{%
	\mathbb{\StrHead{#1}}\mathbf{\StrTail{#1}}
}}

\newcommand{\corr}[1]{R_{\lattice{#1}}}

\newcommand{\sys}[0]{\mathcal{S}}
\newcommand{\dep}[2]{\langle #1 \leadsto #2 \rangle}

\newcommand{\RelaxRel}[3]{\fcn{relax\_rel}_{#1}[#2,#3]}
\newcommand{\Merge}[4]{\fcn{merge}_{#1}[#2,#3 \rightarrow #4]}
\newcommand{\AddDep}[3]{\fcn{add\_dep}_{#1}[#2 \leadsto #3]}

\newcommand{\TightenRel}[3]{\fcn{tighten\_rel}_{#1}[#2,#3]}
\newcommand{\Split}[4]{\fcn{split}_{#1}[#2 \rightarrow #3,#4]}
\newcommand{\RemDep}[3]{\fcn{remove\_dep}_{#1}[#2 \leadsto #3]}

\newcommand{\Apply}[2]{\llbracket #1 \rrbracket(#2)}

\usepackage{amsthm}
\theoremstyle{definition}
\newtheorem{defn}{Definition}[section]

\theoremstyle{plain}
\newtheorem{thm}{Theorem}[section]

\usepackage{enumitem} 

\usepackage{graphicx}
\makeatletter
\providecommand{\bigsqcap}{%
  \mathop{%
    \mathpalette\updown\bigsqcup%
  }%
}
\newcommand*{\updown}[2]{%
  \rotatebox[origin=c]{180}{$#1#2$}%
}
\makeatother

\title{Towards Refinement and Generalization of Reliability Models Based on Component States}
\author{
    \IEEEauthorblockN{Natasha Jarus, Sahra Sedigh Sarvestani, and Ali R. Hurson}\\
    \IEEEauthorblockA{Department of Electrical and Computer Engineering \\
        Missouri University of Science and Technology \\ Rolla, MO 65409, USA\\
        Email: \{jarus, sedighs, hurson\}@mst.edu}
}
\date{Nov. 4, 2019}

\begin{document}

\maketitle

\IEEEpubid{} 
\IEEEpubidadjcol{}
\begin{abstract}
	Complex system design often proceeds in an iterative fashion, starting from a high-level model and adding detail as the design matures.
	This process can be assisted by metamodeling techniques that automate some model manipulations and check for or eliminate modeling mistakes.
	Our work focuses on metamodeling reliability models: we describe generalization and refinement operations for these models.
	Generalization relaxes constraints that may be infeasible or costly to evaluate; refinement adds further detail to produce a model that more closely describes the desired system.
	We define these operations in terms of operations on system constraints.
	To illustrate the proposed method, we relate these constraints to a common Markov chain-based reliability modeling formalism.
\end{abstract}

\section{Introduction}

Designers of critical complex systems---such as autonomous vehicles, power grids, or water distribution networks---must ensure their systems can dependably meet performance requirements.
Dependability encompasses a variety of system metrics that describe the ability of a system to continue to provide service as its components degrade.
Among the most common of these metrics is \emph{reliability}: the probability that a system remains functional up to time $t$.
Reliability takes a binary view of system function: components, and the system, are either functional or failed.
Reliability models based on component states compute a system's reliability as a function of the reliabilities of its components.
This function is determined by the structure of the system---how its components are connected.
For example, a power grid consisting of two transmission lines in parallel is more reliable than the system with the same lines connected in series. 

Complex systems are often designed iteratively.
Requirements are gathered and an initial design is prepared, modeled, and analyzed.
Based on the results, the design is modified to better fit the requirements (or the requirements are modified so the design can better fit them) and the process repeats.
Initial designs and models may be quite general; but they become more detailed as the design progresses.
As the design process can have many iterations, \emph{metamodeling} approaches, which model operations applied to models, are often used to reduce the labor involved, eliminate certain modeling mistakes, and even to help explore the design space.

When modifying a model, we typically want to either add more detail---a new component, a stronger constraint on how that component behaves---or we want to remove a constraint that is unrealistic or would render the design infeasible.
The first action we call \emph{refinement} and the second \emph{generalization}.
Refinement can be used to fill out detail in a high-level model that meets design requirements; generalization can be used to ``back out'' of  a design choice that isn't working.
Both can be used together to explore the design space---refinement asks ``what is the smallest detail that could be added to this model?''; generalization asks ``what happens if this detail is removed?''
It is our goal to make these actions explicit and exact, enabling further analysis and software automation.

\IEEEpubidadjcol{}
In this work, we propose a method for generalization and refinement of Markov Imbeddable Structure (MIS) reliability models where system-level states are identified based on component-level states.
The initial state is one where every component is functional; the terminal state is one where enough components have failed to cause system failure, and intermediate states correspond to the system remaining functional despite some the failure of some of its components.
These models describe a system composed of $n$ components as a Markov chain, encoding each component's reliability and the effect of its failure on other components.
The reliability of the system is then the probability that the system remains functional after taking $n$ steps through the Markov chain.
Our work focuses on MIS models where the states of the Markov chain  are defined by component status (e.g., ``component 3 failed'' or ``only component 2 functional'') and where the component status described by a state remains the same regardless of which component's failure is being considered.
This encompasses the vast majority of MIS models, especially as used in practice; however, it does not encompass certain unusual MIS models, such as models of consecutive-$k$-of-$n$ systems.\footnote{
In short, the transition probability matrices for consecutive-$k$-of-$n$ systems are not upper triangular; for more detail, see \cite[pp.~344--345]{kuo_optimal_2003}.}
These we will address in future work.

When formalizing generalization and refinement, we should consider system properties that are preserved by these operations.
Roughly speaking, if the model $m_r$ is a refinement of a model $m_g$, the constraints imposed on the system by $m_r$ should imply the constraints imposed by $m_g$. 
For example, if $m_r$ requires a component $c$ to have reliability $\ge 0.9$, $m_g$ can require that $c$ have reliability $\ge 0.7$---this constraint is strictly weaker than the constraint of $m_r$.
However, $m_g$ could not require $c$ to have reliability $\ge 0.99$.
In other words, a system meeting the requirements of $m_r$ would provide equal or better reliability than a system meeting $m_g$'s requirements alone.
If $m_r$ refines $m_g$, then $m_g$ generalizes $m_r$, so we can use the same implication relationship to describe both refinement and generalization.
We formally abstract system properties and implication to analyze the soundness of our definitions of generalization and refinement.

Another advantage of describing refinement and generalization in this fashion is that it can be used for model-to-model transformations as shown in our previous work~\cite{jarus_formalizing_2019}.
Provided another formalism represents some of the same system properties, we can relate these MIS models to this formalism in a way that lets us soundly convert between the two.
Thus, the effort required to develop this formalism enables more than the single application this work discusses.

The rest of this paper is as follows.
Section~\ref{sec:background} provides a summary of the theory behind our approach.
System constraints, generalization, and refinement are defined in Section~\ref{sec:properties}.
These operations are connected to MIS models in Section~\ref{sec:mis}.
Finally, related work is surveyed in Section~\ref{sec:related-work} and Section~\ref{sec:conclusion} presents our conclusions.

\section{Background}
\label{sec:background}

The central theory that underlies the work in this paper has been articulated in our previous work~\cite{jarus_formalizing_2019}.
Here we recap the results in terms of the goals of this paper.

Our goal is to relate two domains---a domain of MIS models and a domain of system properties---so that if a certain set of properties describe a given system, the model generated from those properties also describes the system.
Likewise, if a model describes a system, the properties generated from that model also describe the system.
We use this relationship to define generalization and refinement on MIS models based on generalization and refinement of properties.

For our approach, the domains must both be complete lattices $\lattice{L} \triangleq (\set{L}, \sqsubseteq, \bigsqcup, \bigsqcap, \bot, \top)$.
Recall that $\sqsubseteq$ is a partial order relation; for any subset $\set{L'} \subseteq \set{L}$, $\bigsqcup \set{L'}$ is the least upper bound (\emph{join}) and $\bigsqcap \set{L'}$ the greatest lower bound (\emph{meet}) of $\set{L'}$; and $\bot$ and $\top$ are the least and greatest elements of the lattice.
For $\set{L'} = \{l_1, l_2\}$, we write $\bigsqcap \set{L'}$ as $l_1 \sqcap l_2$ and $\bigsqcup \set{L'}$ as $l_1 \sqcup l_2$.

Suppose we have a complete system properties lattice $\lattice{Prop} \triangleq (\set{Prop}, \Rightarrow, \bigvee, \bigwedge, \bot_P, \top_P)$ (see Sec.~\ref{sec:properties})
and a complete MIS model lattice $\lattice{MIS} \triangleq (\set{MIS}, \sqsubseteq, \bigsqcup, \bigsqcap, \bot_M, \top_M)$ (see Sec.~\ref{sec:mis}).
We order both domains by \emph{specificity}.
Intuitively, properties $p_1$ are more specific than properties $p_2$ (i.e., $p_1 \sqsubseteq p_2$) if $p_1$ provides additional information about the system that $p_2$ does not.
Likewise with models: if $m_1 \sqsubseteq m_2$, $m_1$ may offer more detail about the system; for example, $m_1$ may divide a component in $m_2$ into several components with a more complex interrelationship.
The meet of two properties $p_1 \sqcap p_2$ is their logical conjunction; the join $p_1 \sqcup p_2$ is their disjunction.
We will discuss both of these domains in more detail later in the paper.

We use a Galois connection to soundly relate elements of these two domains.
A Galois connection between complete lattices is a pair of functions $\alpha$ and $\gamma$ with properties similar to, but less strict than, those of an order isomorphism.
Informally, Galois connections allow one of the lattices to have ``more detail'' than the other; they are often used in cases where one lattice is an abstraction of the other.

\begin{defn}
	\label{defn:gc}
	A \emph{Galois connection} $(\lattice{P}, \alpha, \gamma, \lattice{M})$ between complete lattices $\lattice{P}$ and $\lattice{M}$ is a pair of functions
	$\alpha : \lattice{P} \rightarrow \lattice{M}$ and $\gamma : \lattice{M} \rightarrow \lattice{P}$ such that
	\begin{enumerate}[label=(\roman*), ref=\ref{defn:gc}.(\roman*)]
		\item\label{defn:gc:i} $\forall p \in \lattice{P}, p \sqsubseteq (\gamma \circ \alpha)(p)$ and
		\item\label{defn:gc:ii} $\forall m \in \lattice{M}, (\alpha \circ \gamma)(m) \sqsubseteq m$.
	\end{enumerate}
\end{defn}
$\alpha$ is called the \emph{abstraction function} (or \emph{abstraction operator}); $\gamma$ is called the \emph{concretization function (operator)}.


Given a Galois connection $(\lattice{Prop}, \alpha, \gamma, \lattice{MIS})$, what do properties \ref{defn:gc:i} and \ref{defn:gc:ii} mean in terms of system properties and MIS models?
Property~\ref{defn:gc:i} states that for every collection of properties $p$, $p \Rightarrow (\gamma \circ \alpha)(p)$: if we abstract a model from $p$, then concretize properties from that model; the result is at worst more general than the properties with which we began.
Likewise, property~\ref{defn:gc:ii} states that for every MIS model $m$, $(\alpha \circ \gamma)(m) \sqsubseteq m$.
Thus, concretizing properties from an MIS model, then abstracting a model from those properties, produces at worst a model more specific than the initial model.
(It is often the case that the $\sqsubseteq$ in \ref{defn:gc:ii} is equality.)

What remains is to relate our domains and the Galois connection between them to a notion of \emph{soundness}.
Soundness is a relative property; whether a model or a collection of properties is sound or not depends on the system being modeled.
Let $\sys \in \set{Sys}$ denote the system we are modeling.
We encode soundness by a relation:
\begin{defn}
	\label{defn:sound-rel}
	A relation $\corr{L} : \set{Sys} \rightarrow \lattice{L}$ between systems and elements of a lattice $\lattice{L}$ is a \emph{soundness relation} if
	\begin{enumerate}[label=(\roman*), ref=\ref{defn:sound-rel}.(\roman*)]
		\item\label{defn:sound-rel:i}  if $\sys\,\corr{L}\,l_1$ and $l_1 \sqsubseteq l_2$, then $\sys\,\corr{L}\,l_2$ and
		\item\label{defn:sound-rel:ii} if $\set{L'} \subseteq \lattice{L}$ and $\forall l \in \set{L'}, \sys\,\corr{L} l$, then $\sys\,\corr{L}\,\bigsqcap \set{L'}$.
	\end{enumerate}
\end{defn}

We suppose that we have a soundness relation $\corr{P} : \set{Sys} \rightarrow \lattice{Prop}$ such that $\sys\,\corr{P}\,p$ if and only if the properties in $p$ describe $\sys$.
Every generalization of a correct collection of properties is sound by property~\ref{defn:sound-rel:i}.
Not every refinement of a collection of properties is necessarily sound---otherwise, every property would be sound for every system.
However, if we know several sound properties, property~\ref{defn:sound-rel:ii} states that they can be refined to a single sound property that implies all known sound properties.

Given the soundness relation $\corr{P}$, we can induce a soundness relation $\corr{M} : \set{Sys} \rightarrow \lattice{MIS}$ by $\sys\,\corr{M}\,m \iff \sys\,\corr{P}\,\gamma(m)$.
Therefore, if properties $p_r$ soundly refine $p_g$, then $\alpha(p_r)$ soundly refines $\alpha(p_g)$.
In short, we need only consider the soundness of refinements in $\lattice{Prop}$; the soundness of our MIS models follows.

\section{Properties}
\label{sec:properties}

Before we describe refinement and generalization of MIS models, we formalize the constraints they place on system design.
The MIS models we consider in this work place three broad constraints on a system: what components are in the system, how reliable each component is, and which components depend on others to remain functional.
The properties domain $\lattice{Prop}$ defines these as a lattice, allowing us to relate these properties to MIS models.

As we will need some way to identify components, let $\set{Comps} \triangleq \{c_1,c_2,\dotsc\}$ be the set of all possible component names.

Each element $p \in \lattice{Prop}$ is a triplet $p = (\set{C}, \fcn{R}, \set{D})$ where
\begin{itemize}
	\item $\set{C} \subseteq \set{Comps}$ is the finite set of names of components in the system (e.g., $\{c_1, c_2, c_3\}$);
	\item $\fcn{R} : \set{C} \rightarrow [0, 1]$ is a function that specifies a lower bound for the reliability of each component: if the reliability of $c$ is $p$, then $\fcn{R}(c) \leq p$; and
	\item $\set{D} \subseteq \set{Deps}$ is the finite set of component dependencies, as described in the next section.
\end{itemize}
For example, a system consisting of two 90\% reliable power lines in parallel where the failure of one causes the other to become overloaded and thus fail as well would be described by the properties
$(\set{C} = \{c_1, c_2\}, \fcn{R}(c_1) = \fcn{R}(c_2) = 0.9, \set{D} = \{\dep{c_1}{c_2, \sys}, \dep{c_2}{c_1, \sys}\})$.

\subsection{Dependencies}

Component dependencies (elements of $\set{Deps}$)  are represented by the relation $\dep{\_}{\_} : \powerset{\set{C}} \rightarrow \powerset{\set{C} \cup \{\sys\}}$.\footnote{
$\powerset{\set{S}}$ denotes the set of subsets (``powerset'') of the set $\set{S}$.}
The statement $\dep{\cdots_1}{\cdots_2}$ means ``the failure of the components in the set $\cdots_1$ immediately leads to the failure of the components in $\cdots_2$''.
Should $\sys$ appear in $\cdots_2$, the system also fails as a result of the components of $\cdots_1$ failing.
The components on the left side ($\cdots_1$) are referred to as \emph{causes} and the components on the right ($\cdots_2$) as \emph{effects}.

These dependencies correspond to state transitions.
Suppose we have a system with components $\set{C} = \{c_1, c_2, c_3\}$.
We can represent the state of the components as three-bit strings: \state{111} corresponds to the system state where all components are functional, \state{101} corresponds to the state where $c_2$ has failed, etc.
A dependency $\dep{c_1}{\emptyset}$ corresponds to a transition from \state{111} to \state{011} when $c_1$ fails---the failure of $c_1$ does not influence the functionality of other components in the system.
Likewise, a dependency $\dep{c_1, c_2}{c_3,\sys}$ corresponds to transitions from \state{101} to \state{000} when $c_1$ fails and from \state{011} to \state{000} when $c_2$ fails; furthermore, in state \state{000} the system is considered failed.
Sec.~\ref{sec:mis} formalizes this correspondence.

As there are a number of ways to write dependencies, we place some constraints on them to ensure the constraints on the system are consistent with how components fail and fully cover all cases of system behavior.
These constraints are split into \emph{equivalences} and \emph{well-formedness (WF) properties}.

\subsubsection{Equivalences}
The first equivalence rule states that if a component appears on both sides of $\leadsto$, we can remove it from the right side.
The failure of any component trivially causes that component to fail; this rule states that we need not write this fact explicitly:\footnote{%
A note on notation: $c\cdots_1$ refers to a set containing the component $c$ and the components of the set $\cdots_1$.
}
\begin{equation}
	\tag{Tautology}
	\dep{c\cdots_1}{c\cdots_2} \equiv \dep{c\cdots_1}{\cdots_2}.
	\label{eqn:taut}
\end{equation}

The remaining two equivalences are between sets of dependencies, rather than between two individual dependencies.
If we have two dependencies with the same cause but different effects, we can produce one dependency that represents both by taking the union of their effects:
\begin{equation}
	\tag{Union}
	\left\{
		\begin{gathered}
			\dep{\cdots_1}{\cdots_2} \\
			\dep{\cdots_1}{\cdots_3}
		\end{gathered}
		\right\}
	\equiv \left\{\dep{\cdots_1}{\cdots_2\cdots_3}\right\}.
	\label{eqn:union}
\end{equation}

Finally, a dependency with no causes cannot occur:
\begin{equation}
	\tag{Inaction}
	\left\{\dep{\emptyset}{\cdots}\right\} \equiv \emptyset.
	\label{eqn:inaction}
\end{equation}

\subsubsection{Well-formedness Properties}

The WF properties describe a system-level view of dependencies: what dependencies need to be present in $\set{D}$ to make a consistent set of system constraints.
First, every component must have a dependency where it is the sole cause of failure (although the effect may be the empty set). 
These correspond to transitions from the initial \state{1\cdots1} state:
\begin{equation}
	\tag{Initiality}
	\forall c \in \set{C}, \exists \dep{c}{\cdots} \in \set{D}.
	\label{eqn:initial}
\end{equation}

In addition, at least one sequence of failures must lead to the system failing (otherwise, the system's reliability would be 1 and there would be nothing to model):
\begin{equation}
	\tag{Termination}
	\exists \dep{\cdots_1}{\sys \cdots_2} \in \set{D}.
	\label{eqn:termination}
\end{equation}

Finally, components cannot recover as a result of the failure of other components.
Thus, if components $\cdots_1$ cause components $\cdots_2$ to fail, any other dependency where $\cdots_1$ have failed must also have $\cdots_2$ failed.
\begin{equation}
	\tag{Monotonicity}
	\begin{split}
		&\forall \dep{\cdots_1}{\cdots_2} \in \set{D},\\
		&\forall \dep{\cdots_1\cdots_3}{\cdots_4} \in \set{D},\\
		&\cdots_2 \subseteq \cdots_3 \cup \cdots_4.
	\end{split}
	\label{eqn:monotonicity}
\end{equation}

For instance, if we have $\dep{c_1}{c_2}$, \ref{eqn:monotonicity} would permit the dependencies $\dep{c_1,c_3}{c_2}$ and $\dep{c_1,c_2}{c_3}$ but forbid $\dep{c_1,c_3}{\emptyset}$, as $c_2$ must always fail when $c_1$ fails.

\subsubsection{Examples}

Before addressing generalization and refinement of properties, we demonstrate a few examples of how dependencies are used to specify system behavior.
First, consider the dependencies in the earlier parallel-component example: $\set{D} = \{\dep{c_1}{c_2, \sys}, \dep{c_2}{c_1, \sys}\}$.
In this system, the failure of component $c_1$ leads to the failure of $c_2$ and system failure, and vice versa for $c_2$.
This system has two states, \state{11} and \state{00}; the failure of either component causes a transition from the first to the second.

By contrast, a parallel-component system where the two components are independent would be specified by $\set{D} = \{\dep{c_1}{\emptyset},\dep{c_2}{\emptyset},\dep{c_1,c_2}{\sys}\}$.
This system has all four possible states and all valid transitions between states.

A system with two components in series produces a more interesting ``failed'' state.
These components are independent, as one failing does not cause the other to fail, but both need to be functional for the system to function: $\set{D} = \{\dep{c_1}{\sys},\dep{c_2}{\sys}\}$.
This system also has two states: the initial state \state{11} and the failed superstate \twostate{01}{10}.\footnote{%
MIS modeling requires a single ``failed'' system (super)state; we leave unification of functional states into superstates for future work.}
Once the system has failed, we are no longer interested in its behavior; thus, for this system, we consider \state{00} unreachable.

\subsection{Generalization}

Now that we have described the elements of $\lattice{Prop}$, we can describe how to generalize them.
The goal of generalizing an element of $\lattice{Prop}$ is to produce an element of $\lattice{Prop}$ that relaxes the constraints of the first element but does not contradict it.
Understanding how constraints can be generalized allows us to order $\lattice{Prop}$ by generalization.

\subsubsection{One-step generalizations of dependencies}

For a given reliability model, one way to generalize dependencies is to lower the constraint on a component's reliability: a more reliable component can always be substituted for a less reliable one.
We can relax the reliability of a component, $c$, to a lower constraint $r < \fcn{R}(c)$ by
\begin{align}
	& \RelaxRel{(\set{C}, \fcn{R}, \set{D})}{\_}{\_}: \set{C} \rightarrow [0,1] \rightarrow \lattice{Prop} \notag \\
	& \RelaxRel{(\set{C}, \fcn{R}, \set{D})}{c}{r} \triangleq (\set{C}, \fcn{R'}, \set{D})
\end{align} 
where
\begin{equation}
	\tag{1.1}
	\fcn{R'}(c') \triangleq \begin{cases}
		r & \text{if } c = c' \\
		\fcn{R}(c) & \text{otherwise.}
	\end{cases}
\end{equation}

The other means of generalizing system constraints is to generalize component dependencies.
We begin by considering the smallest actions we can take that generalize system dependencies while maintaining the WF properties.
There are two possible operations: merging two components and adding a new dependency $\dep{\cdots}{c}$ among existing components.
Both of these operations take one element of $\lattice{Prop}$ and infer another.

Two distinct components $c_1$ and $c_2$ can be merged into a single component $c_m$ (where the name $c_m$ does not already appear in $\set{C} \setminus \{c_1, c_2\}$)
by replacing every instance of $c_1$ and $c_2$ with $c_m$:
\begin{align}
	&\Merge{(\set{C}, \fcn{R}, \set{D})}{\_}{\_}{\_}: \set{C} \rightarrow \set{C} \rightarrow \set{Comps} \rightarrow \lattice{Prop} \notag \\
	&\Merge{(\set{C}, \fcn{R}, \set{D})}{c_1}{c_2}{c_m} \triangleq (\set{C'}, \fcn{R'}, \set{D'})
\end{align} 
where
\begin{align}
	\tag{2.1}
	\set{C'} \triangleq{}& \{c_m\} \cup \set{C} \setminus \{c_1, c_2\} \\
	\tag{2.2}
	\fcn{R'}(c) \triangleq{}&
	\begin{cases}
		\min(\fcn{R}(c_1), \fcn{R}(c_2)) & \text{if } c = c_m, \\
		\fcn{R}(c) & \text{otherwise}.
	\end{cases}
	\\
	\tag{2.3}
	\set{D'} \triangleq{}& \{ \dep{m(\set{c})}{m(\set{e})} \mid \dep{\set{c}}{\set{e}} \in \set{D}\} \\
	\tag{2.4}
	m(\set{c}) \triangleq{}&
	\begin{cases}
		\{c_m\} \cup \set{c} \setminus \{c_1, c_2\} & \text{if } c_1 \in \set{c} \vee c_2 \in \set{c}, \\
		\set{c} & \text{otherwise}.
	\end{cases}
\end{align}

When defining a generalization, we should ensure that it only relaxes constraints.
Thus, when choosing the reliability bound $\fcn{R'}(c_m)$ of the merged component, we must pick the least restrictive choice $\min(\fcn{R}(c_1), \fcn{R}(c_2))$.
Effectively, this choice performs two generalizations: first, we relax the tighter of the reliability bounds of $c_1$ and $c_2$ by setting $\fcn{R}(c_1) = \fcn{R}(c_2)$,
then we merge $c_1$ and $c_2$ into one component.


The other possible generalization is adding a dependency among existing components.
This may seem counterintuitive; however, it is a stronger claim to say that a component is independent of another---the fewer dependencies a system has, the more reliable it is.
Adding a dependency from a nonempty set of components $\set{c}$ to a component $e \notin \set{c}$ means that whenever the components in $\set{c}$ cause a failure, $e$ is amongst the effects.
As all the components in $\set{c}$ and $e$ are in $\set{C}$ already, we need only modify the dependencies:
\begin{align}
	& \AddDep{(\set{C}, \fcn{R}, \set{D})}{\_}{\_} : \powerset{\set{C}} \rightarrow \set{C} \rightarrow \lattice{Prop} \notag \\
	& \AddDep{(\set{C}, \fcn{R}, \set{D})}{\set{c}}{e} \triangleq (\set{C}, \fcn{R}, \set{D'}) \label{eqn:add-dep}
\end{align} 
where
\begin{align}
	\tag{3.1}
	\set{D'} \triangleq{}& \{a(\dep{\set{c'}}{\set{e'}}) \mid \dep{\set{c'}}{\set{e'}} \in \set{D}\} \\
										 &\cup \{\dep{\set{c}}{\set{u} \cup \{e\}}\} \notag\\
  \tag{3.2}\label{eqn:add-dep:a}
	a(\dep{\set{c'}}{\set{e'}}) \triangleq{}&
	\begin{cases}
			\dep{\set{c'} \setminus \{e\}}{\set{e'} \cup \{e\}} & \text{if } \set{c} \subseteq \set{c'}, \\ 
			\dep{\set{c'}}{\set{e'}}& \text{otherwise}.
		\end{cases}\\
	\tag{3.3}
	\set{u} \triangleq{}& \bigcup \{\set{e'} \mid \dep{\set{c'}}{\set{e'}} \in \set{D} \text{ where } \set{c'} \subseteq \set{c}\}
\end{align}


For an example of the effect of generalization operations on a system, consider a system with three independent components:
\begin{equation*}
	\begin{split}
		p ={} &(\set{C} = \{c_1, c_2, c_3\}, \fcn{R}(\_) = 0.9, \set{D} = \{\\
				& \quad \begin{split}
							&\dep{c_1}{\emptyset},\dep{c_2}{\emptyset},\dep{c_3}{\emptyset},\\
							&\dep{c_1,c_2,c_3}{\sys}
				\end{split}\\
				&\})
	\end{split}
\end{equation*}

Introducing a dependency $\dep{c_1,c_2}{c_3}$ results in the following system:
\begin{equation*}
	\begin{split}
		p' ={} & \AddDep{p}{c_1,c_2}{c_3}\\
		   ={} & (\set{C'} = \{c_1, c_2, c_3\}, \fcn{R'}(\_) = 0.9, \set{D'} = \{\\
				& \quad \begin{split}
							&\dep{c_1}{\emptyset},\dep{c_2}{\emptyset},\dep{c_3}{\emptyset},\\
							&\negmedspace\left.
								\begin{aligned}
									&\dep{c_1,c_2}{c_3}^\dag,\\
									&\dep{c_1,c_2}{c_3,\sys}^\ddag
								\end{aligned}	
								\right\}
								\equiv \dep{c_1,c_2}{c_3,\sys}
				\end{split}\\
				&\})
	\end{split}
\end{equation*}
Of note: the dependency marked $^\dag$ is the new dependency added by $\fcn{add\_dep}$ and the dependency marked $^\ddag$ is the result of the first substitution rule in (\ref{eqn:add-dep:a}).
Both rules reduce to one via the \ref{eqn:union} property.



\subsubsection{Multi-step generalization of dependencies}

The example of the previous section illustrates the process by which successive generalization steps are applied to system properties.
To describe this more formally, let $\set{G}$ be the set of all generalization operations and $\set{G}^*$ be the set of finite sequences of elements of $\set{G}$.
We define the act of applying a sequence of generalizations to an element of properties, $\Apply{\_}{\_}: \set{G}^* \rightarrow \lattice{Prop} \rightarrow \lattice{Prop}$, by
\begin{equation}
	\Apply{g}{p} \triangleq \begin{cases}
		p & \text{if } g = () \\
		\llbracket gs \rrbracket(g'_p) & \text{if } g = (g', gs).
	\end{cases}
\end{equation}

With the ability to apply a sequence of generalizations, we now turn to the task of ordering elements of $\lattice{Prop}$.
%

\subsubsection{Generalization as a partial order}

To form a partial order on  $\lattice{Prop}$ using these generalization operations, we say that if $p_g$ generalizes $p_r$, there exists some sequence of generalizations that witnesses that fact:
\begin{defn}
	$p_g = \in \lattice{Prop}$ \emph{generalizes} $p_r \in \lattice{Prop}$, written $p_r \sqsubseteq p_g$, if
	$\exists g \in \set{G}^*, \Apply{g}{p_r} = p_g$.
\end{defn}

\begin{thm}
	$\sqsubseteq$ forms a partial order on $\lattice{Prop}$.
\end{thm}

\subsection{Refinement}

In addition to generalization of constraints, we are interested in refining them: adding new constraints or increasing the strictness of existing ones.
Refinements are dual to generalizations, so for each generalization we expect a corresponding refinement.

\subsubsection{One-step Refinements}

Corresponding to $\fcn{relax\_rel}$ we have $\fcn{tighten\_rel}$ which raises the bound on the reliability of component $c$ to a higher constraint $r > \fcn{R}(c)$:
\begin{align}
	& \TightenRel{(\set{C}, \fcn{R}, \set{D})}{\_}{\_} : \set{C} \rightarrow [0,1] \rightarrow \lattice{Prop} \notag \\
	& \TightenRel{(\set{C}, \fcn{R}, \set{D})}{c}{r} \triangleq (\set{C}, \fcn{R'}, \set{D})
\end{align}
where
\begin{equation}
	\tag{5.1}
	\fcn{R'}(c') \triangleq \begin{cases}
		r & \text{if } c = c' \\
		\fcn{R}(c') & \text{otherwise}
	\end{cases}
\end{equation}

To undo a \fcn{merge}, we split one component, $c_m$, into two, $c_1$ and $c_2$ (where $c_1, c_2 \notin \set{C} \setminus \{c\}$).
When splitting two components, we make each fully dependent on the other, as that is the most general set of constraints we can generate.
In other words, the result of $\Split{p}{c_m}{c_1}{c_2}$ is the maximal element of the set $\{q \in \lattice{Prop} \mid p = \Merge{q}{c_1}{c_2}{c_m}\}$.
\begin{align}
	& \Split{(\set{C}, \fcn{R}, \set{D})}{\_}{\_}{\_} : \set{C} \rightarrow \set{Comps} \rightarrow \set{Comps} \rightarrow \lattice{Prop} \notag \\
	& \Split{(\set{C}, \fcn{R}, \set{D})}{c_m}{c_1}{c_2} \triangleq (\set{C'}, \fcn{R'}, \set{D'})
\end{align}
where
\begin{align}
	\tag{6.1}
	\set{C'} \triangleq{}& \{c_1,c_2\} \cup \set{C} \setminus \{c_m\} \\
	\tag{6.2}
	\fcn{R'}(c) \triangleq{}&
		\begin{cases}
			\fcn{R}(c_m) & \text{if } c = c_1 \vee c = c_2, \\
			\fcn{R}(c) & \text{otherwise}.
		\end{cases}
		\\
	\tag{6.3}
	\set{D'} \triangleq{}& \bigcup\{ s(\dep{\set{c}}{\set{e}}) \mid \dep{\set{c}}{\set{e}} \in \set{D}\} \\
	\tag{6.4}
	s(\dep{\set{c}}{\set{e}}) \triangleq{}&
		\begin{cases}
			\left\{\begin{aligned}
					& \dep{\{c_1,c_2\} \cup \set{c'}}{\set{e}}\\
					& \dep{\{c_1\} \cup \set{c'}}{\set{e} \cup \{c_2\}}\\
					& \dep{\{c_2\} \cup \set{c'}}{\set{e} \cup \{c_1\}}
		\end{aligned}\right\} & \text{if } c_m \in \set{c} \\
		\left\{\begin{aligned}
				& \dep{\set{c}}{\set{e'} \cup \{c_1,c_2\}} \\
				& \dep{\set{c}}{\set{e'} \cup \{c_1\}} \\
				& \dep{\set{c}}{\set{e'} \cup \{c_2\}}
		\end{aligned}\right\} & \text{if } c_m \in \set{e} \\
			\{\dep{\set{c}}{\set{e}}\} & \text{otherwise}.
		\end{cases}\\
	\tag{6.5}
	\set{c'} \triangleq{}& \set{c} \setminus \{c_m\} \\
	\tag{6.6}
	\set{e'} \triangleq{}& \set{e} \setminus \{c_m\}
\end{align}


Finally, $\fcn{remove\_dep}$ corresponds to undoing an $\fcn{add\_dep}$ operation.
Adding a dependency $\dep{\cdots_1}{e}$ states that $e$ depends on all of $\cdots_1$ and therefore every dependency containing $\cdots_1$ is rewritten to preserve \ref{eqn:monotonicity}.
Removing a dependency $\dep{\cdots_1}{e}$ states that $e$ is \emph{independent} of all components in $\cdots_1$, so every dependency whose causes are contained in $\cdots_1$ is rewritten.
\begin{align}
	& \RemDep{(\set{C}, \fcn{R}, \set{D})}{\_}{\_} : \powerset{\set{C}} \rightarrow \set{C} \rightarrow \lattice{Prop} \notag \\
	& \RemDep{(\set{C}, \fcn{R}, \set{D})}{\set{c}}{e} \triangleq (\set{C}, \fcn{R}, \set{D'})
\end{align}
where
\begin{align}
	\tag{7.1}
	\set{D'} \triangleq{}& \{r(\dep{\set{c'}}{\set{e'}}) \mid \dep{\set{c'}}{\set{e'}} \in \set{D}\} \\
	\tag{7.2}
	r(\dep{\set{c'}}{\set{e'}}) \triangleq{}&
	\begin{cases}
			\dep{\set{c'}}{\set{e'} \setminus \{e\}} & \text{if } \set{c'} \subseteq \set{c}, \\
			\dep{\set{c'}}{\set{e'}}& \text{otherwise}.
		\end{cases}
\end{align}


\subsubsection{Multi-step Refinements}

As with generalizations, let $\set{R}$ be the set of all refinement operations and $\set{R}^*$ be the set of all sequences of refinements.
We abuse notation slightly to define application of a sequence of refinements using the same notation: for $rs \in \set{R}^*$, $\Apply{rs}{p}$ is the result of applying that sequence of refinements to some system properties $p$.

\subsubsection{Refinement as the dual of generalization}

Each generalization operation and its corresponding refinement are not necessarily inverses, as most generalization operations map several elements of $\lattice{Prop}$ to the same more general system (i.e., they are not injective).
Thus, we do not have that $\forall g \in \set{G}$, if $q = \Apply{g}{p}$ then $\exists r \in \set{R}, p = \Apply{r}{q}$.
However, we can show the opposite: if $q = \Apply{r}{p}$, then $p$ \emph{covers} $q$: there is no $r$ such that $q \sqsubset r \sqsubset p$.

Furthermore, the refinement operations form a dual order to the order defined by generalization:
\begin{thm}
	$\forall p_r,p_g \in \lattice{Prop}$, $p_r \sqsubseteq p_g$ if and only if $\exists rs \in \set{R}^*, p_r = \Apply{rs}{p_g}$.
\end{thm}
As such, $p_r$ \emph{refines} $p_g$ if $p_g \sqsupseteq p_r$, or, equivalently, $p_r \sqsubseteq p_g$.

\subsection{The Properties Lattice}

To be able to use a Galois connection to relate our notions of generalization and refinement to MIS models, we must define $\lattice{Prop}$ as a lattice.
As such, we need to define top and bottom elements of $\lattice{Prop}$, least upper bounds (or \emph{joins}), and greatest lower bounds (\emph{meets}).\footnote{%
Discussion of meets and joins is omitted for lack of space.}

The top element of $\lattice{Prop}$ is the one-element system with unconstrained component reliability:
\begin{equation}
	\top \triangleq (\{c\}, \fcn{R}(c) = 0, \{\dep{c}{\sys}\}).
\end{equation}
Any other one-element system constrains component reliability and thus can be generalized to $\top$ by $\fcn{relax\_rel}$.
Removing the one dependency results in a system that does not meet the WF properties, and no further dependencies can be added without adding another component.
Finally, given $p \in \lattice{Prop}$, we can show $p \sqsubseteq \top$ by repeatedly merging components in $p$ until the result has one component, then relaxing that component's reliability bound, if necessary.

The bottom element of $\lattice{Prop}$ is a special element which corresponds to an ``overdetermined'' system---one where the constraints are contradictory.
We do not concern ourselves with its representation, but simply define it as the element $\bot \in \lattice{Prop}$ such that $\forall p, \bot \sqsubseteq p$.



\section{MIS Models}
\label{sec:mis}

Markov Imbeddable Structure models are one approach to deriving a system's reliability from the reliability of its components.
These models consist of states and transitions between states caused by the failure of components.
The reliability of the system is determined by computing the probability of the system not reaching the ``failed'' state after considering the effect of each component.

This paper considers MIS models where the states are defined by the components functional in that state; e.g., \state{1101} corresponds to the state of a 4-component system where components 1, 2, and 4 are functional and component 3 has failed.
Components cannot repair themselves, so every transition is either from one state to that same state or from one state to a state with more failed components.
The \emph{failed} state is absorbing---once the system fails, we are no longer interested in its behavior.

These transitions are usually represented in the form of transition probability matrices (TPMs) $T_i$, one for each component.
As the system always starts in the fully functional state, the initial state probability vector is $\Pi_0 \triangleq [1,0,\dotsc]$.
Another vector $u \triangleq [1,\dotsc, 0]$ defines which states are considered functional.
The system reliability is given by the product of the initial state probabilities, the TPMs, and the $u$ vector:
\begin{equation}
	\fcn{R}(\sys) \triangleq \Pi_0^T * T_1 * T_2 * \dotsb * T_n * u
\end{equation}

As an example, consider the system with two components in series where $\fcn{R}(c_1) = \fcn{R}(c_2) = p = 1 - q$.
The TPM for both components is given by
\begin{equation*}
	T_1 = T_2 = \begin{pmatrix}
			p & q  \\
			0 & 1
		\end{pmatrix}
\end{equation*}
and the resulting system reliability is
\begin{equation*}
	\fcn{R}(\sys) = \Pi_0^T * T_1 * T_2 * u = p^2
\end{equation*}
%
%


\subsection{Abstraction and Concretization}

To apply our formalization of refinement and generalization to MIS models, we need to connect our properties domain $\lattice{Prop}$ to MIS models.
We achieve this by an \emph{abstraction} operator which converts system constraints to MIS models and a \emph{concretization} operator which derives constraints from MIS models.

To abstract an MIS model from $(\set{C}, \fcn{R}, \set{D}) \in \lattice{Prop}$, for each $c_i \in \set{C}$ let $p_i = 1 - q_i = \fcn{R}(c_i)$ be its reliability and let $T_i$ be its TPM.
Let $n = |\set{C}|$ be the number of components in the system.
Then, begin with the initial fully-functional state \state{1\cdots1}.
For each dependency $\dep{c_i}{\set{e}} \in \set{D}$, insert a transition from \state{1\cdots1} to \state{1\cdots1} with probability $p_i$ in $T_i$
and a transition from \state{1\cdots1} to the state where all components except $c_i$ and those in $\set{e}$ are functional with probability $q_i$ in $T_i$.
If $\sys \in \set{e}$, then mark that state as ``failed''.
For each non-``failed'' state added in the previous step, let $\set{s}$ be the components functional in that state and let $\set{f} = \set{C} \setminus \set{s}$ be the set of failed components.
For each component $c_i \in \set{s}$, select the dependency $\dep{\set{c}}{\set{e}} \in \set{D}$ where $c_i \in \set{c}$ and $\set{c}$ is the largest set such that $\set{c} \subset \set{f}$.
Insert transitions from \state{\set{s}} to \state{\set{s}} with probability $p_i$ and from \state{\set{s}} to \state{\set{s}\setminus \set{e}} with probability $q_i$ into $T_i$.
For each component $c_i \in \set{f}$, insert a transition from \state{\set{s}} to \state{\set{s}} with probability $1$ into $T_i$.
Repeat this step until there are no more non-failed states to consider.


Concretizing properties from an MIS model proceeds in an analogous fashion.
For each $T_i$ create a component $c_i$ and set $\fcn{R}(c_i) = p_i$.
For each $c_i$, first let $\set{s'}$ be the set of components functional after $c_i$ fails from the initial \state{1\cdots1} state
and add a dependency $\dep{c_i}{\set{C} \setminus \set{s'}}$ to $\set{D}$.
Then consider all transitions in $T_i$ from state \state{\set{s}} to state \state{\set{s'}} where $\set{s'} \subset \set{s}$.
Let $\set{f} \triangleq \set{s} \setminus \set{s'} \setminus \{c_i\}$ be the set of components that also fail as a result of the failure of $c_i$.
Take $\dep{\set{c}}{\set{e}} \in \set{D}$ where $c_i \in \set{c}$ and $\set{c}$ is the largest set such that $\set{c} \subset (\set{C} \setminus \set{s})$.
If $\set{e} \ne \set{f}$, add a dependency $\dep{\set{C} \setminus \set{s} \setminus \{c_i\}}{\set{f}}$.



\subsection{Examples}

As an example of the power of this approach, let us refine a 2-of-3 system from $\top$.
Our starting system is
\begin{equation*}
	\top = (\{c_1\}, \fcn{R}(c_1) = 0, \{\dep{c_1}{\sys}\}).
\end{equation*}
If we refine $c_1$'s reliability to $p$ by $s_1 = \TightenRel{\top}{c_1}{p}$, the resulting system has reliability $\fcn{R}(\sys) = p$.
%

First, we create another component via $s_2 = \Split{s_1}{c_1}{c_1}{c_2}$, we get the following system:
\begin{equation*}
	\begin{split}
		s_2 ={}& (\{c_1, c_2\}, \fcn{R}(c_1) = \fcn{R}(c_2) = p, \{\\
				& \qquad \dep{c_1}{c_2,\sys},\dep{c_2}{c_1,\sys}\\
				& \qquad \dep{c_1,c_2}{\sys}\\
				& \})
	\end{split}
\end{equation*}
%
This gives $\fcn{R}(\sys) = p^2$ as we now take two steps through the Markov chain.

We can avoid adding excessive dependencies later by removing two, making $c_1$ independent: $s_3 = \RemDep{s_2}{c_1}{c_2,\sys}$.
\begin{equation*}
	\begin{split}
		s_3 ={}& (\{c_1, c_2\}, \fcn{R}(c_1) = \fcn{R}(c_2) = p, \{\\
				& \qquad \dep{c_1}{\emptyset}, \dep{c_2}{c_1,\sys}\\
				& \qquad \dep{c_1,c_2}{\sys}\\
				& \})
	\end{split}
\end{equation*}
Removing these dependencies adds a new state to the Markov chain:
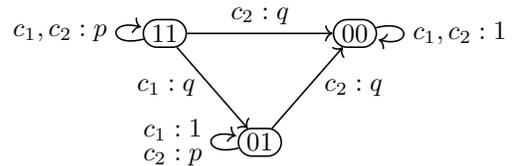
\begin{center}
	\begin{tikzpicture}[->,auto,node distance=1.5cm,semithick,every text node part/.style={align=center}]
	\node[shape=rounded rectangle,draw,inner sep=2pt] (s11) {$11$};
	\node[shape=rounded rectangle,draw,inner sep=2pt, below right=of s11] (s01) {$01$};
	\node[shape=rounded rectangle,draw,inner sep=2pt, above right=of s01] (sf) {$00$};

	\path
	(s11) edge [loop left] node {$c_1, c_2: p$} (s11)
	(s11) edge  node {$c_2: q$} (sf)
	(s11) edge  node [below, left=0.1cm] {$c_1: q$} (s01)
	(s01) edge [loop left] node {$c_1 : 1$ \\[-3pt] $c_2:p$} (s01)
	(s01) edge  node [below, right=0.1cm] {$c_2: q$} (sf)
	(sf) edge [loop right] node {$c_1, c_2: 1$} (sf)
	;
\end{tikzpicture}
\end{center}
This gives $\fcn{R}(\sys) = p^2 + pq$---either both components remain functional, or $c_1$ fails and $c_2$ remains functional.

Next, we introduce $c_3$ by $s_4 = \Split{s_3}{c_2}{c_2}{c_3}$.
\begin{equation*}
	\begin{split}
		s_4 ={}& (\{c_1, c_2, c_3\}, \fcn{R}(c_1) = \fcn{R}(c_2) = \fcn{R}(c_3) = p, \{\\
				& \qquad \dep{c_1}{\emptyset}, \dep{c_2}{c_1,c_3,\sys}, \dep{c_3}{c_1,c_2,\sys}\\
				& \qquad \dep{c_1,c_2}{\sys}, \dep{c_1,c_3}{c_2,\sys}, \dep{c_2,c_3}{c_1,\sys}\\
				& \})
	\end{split}
\end{equation*}
The Markov chain is similar to the one abstracted from $s_3$, but $c_3$ adds its own transition probabilities.
%
This gives $\fcn{R}(\sys) = p^3 + p^2q$---either all components remain functional, or $c_1$ fails and $c_2$ and $c_3$ remain functional.

Finally, we arrive at the desired 2-of-3 system by removing unneeded dependencies: $s_5 = \RemDep{s_4}{c_2}{c_1,c_3,\sys}$ and $s_6 = \RemDep{s_5}{c_3}{c_1,c_2,\sys}$.
\begin{equation*}
	\begin{split}
		s_6 ={}& (\{c_1, c_2, c_3\}, \fcn{R}(c_1) = \fcn{R}(c_2) = \fcn{R}(c_3) = p, \{\\
				& \qquad \dep{c_1}{\emptyset}, \dep{c_2}{\emptyset}, \dep{c_3}{\emptyset}\\
				& \qquad \dep{c_1,c_2}{\sys},\dep{c_1,c_3}{c_2,\sys}, \dep{c_2,c_3}{c_1,\sys}\\
				& \})
	\end{split}
\end{equation*}
The abstracted Markov chain has two new states:
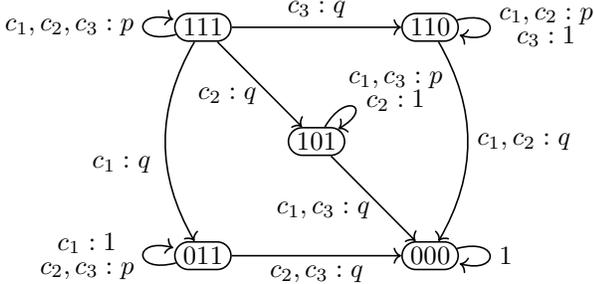
\begin{center}
	\begin{tikzpicture}[->,auto,node distance=1.6cm,semithick,every text node part/.style={align=center}]
	\node[shape=rounded rectangle,draw,inner sep=2pt] (s101) {$101$};
	\node[shape=rounded rectangle,draw,inner sep=2pt, above left=of s101] (s111) {$111$};
	\node[shape=rounded rectangle,draw,inner sep=2pt, above right=of s101] (s110) {$110$};
	\node[shape=rounded rectangle,draw,inner sep=2pt, below left=of s101] (s011) {$011$};
	\node[shape=rounded rectangle,draw,inner sep=2pt, below right=of s101] (sf) {$000$};

	\path
	(s111) edge [loop left] node {$c_1, c_2, c_3 : p$} (s111)
	(s111) edge [bend right] node [below left=0.1cm] {$c_1:q$} (s011)
	(s111) edge  node [below left, inner sep=1pt] {$c_2:q$} (s101)
	(s111) edge  node {$c_3:q$} (s110)
	(s110) edge [loop right] node {$c_1, c_2:p$ \\[-3pt] $c_3:1$} (s110)
	(s110) edge [bend left] node {$c_1, c_2:q$} (sf)
	(s101) edge [out=60, in=30, looseness=8] node [inner sep=-2pt] {$c_1, c_3 : p$ \\[-3pt] $c_2:1$} (s101)
	(s101) edge  node [below left, inner sep=1pt] {$c_1, c_3:q$} (sf)
	(s011) edge [loop left] node {$c_1: 1$ \\[-3pt] $c_2, c_3:p$} (s011)
	(s011) edge  node [below] {$c_2, c_3:q$} (sf)
	(sf) edge [loop right] node {$1$} (sf)
	;
\end{tikzpicture}
\end{center}
This gives $\fcn{R}(\sys) = p^3 + 3p^2q$---either all components remain functional, or only one fails.

\section{Related Work}
\label{sec:related-work}

Markov chains form the theoretical basis for numerous system reliability analyses.
Of particular relevance to this work are two applications of MIS modeling to smart grids---power grids augmented with cyber monitoring and control capabilities to improve their dependability~\cite{albasrawi_analysis_2014,marashi_consideration_2018}.
These studies demonstrate how MIS modeling can be applied to real-world systems to capture system reliability and component interdependencies.

Refinement of specifications for software programs has been studied extensively; see~\cite{morgan_programming_1990} for an introduction and~\cite{gulwani_program_2017} for a recent survey of the literature.
The essence of program specification and refinement is augmenting a programming language with a specification language.
Thus, programs become specifications that are executable.
To derive programs from non-executable specifications, a refinement relation is defined and various refinements of specifications are developed.
This allows one to start with a high-level specification of a program's behavior and derive, through repeated refinement, an executable program whose specification refines the initial specification.

Research on refinement of Markov chains has taken two forms.
The first focuses on Interval Markov Chains (IMCs) and their extension, Constraint Markov Chains (CMCs)~\cite{caillaud_compositional_2010,delahaye_consistency_2012}.
In these formalisms, transition probabilities are not given exactly, but are bounded within an interval or given by algebraic constraints, respectively.
As each IMC or CMC corresponds to a collection of Markov chains that satisfy the requirements given, it is possible to define refinement directly in terms of these formalisms,
rather than using a separate ``system constraints'' formalism, as we do.
Each system specification can be written as an IMC or CMC and then refined into a complete system model via refinement and conjunction operations.

The second approach uses counterexample generation to validate Markov chain abstractions used in model checking~\cite{chadha_counterexample-guided_2010}.
Starting with a coarse approximation of the original Markov chain, model checking is performed until a counterexample is found.
This counterexample is checked against the original specification; if the counterexample does not hold, the approximate system is refined so the counterexample no longer holds.
This process repeats until a genuine counterexample is found (one that holds for the original specification) or the model checking algorithm cannot find a counterexample.
A related work~\cite{kattenbelt_game-based_2010} bounds the uncertainty introduced by this approach to state-space reduction by separately modeling the uncertainty present in the model and the uncertainty added through abstraction.

\section{Conclusion}
\label{sec:conclusion}

In this paper,  we have proposed and demonstrated an approach to refinement and generalization of MIS reliability models.
Key to this approach is a system constraints domain, which captures the behavior of a system in an abstract, easily manipulated fashion.
These constraints describe the components of the system, their reliability, and dependencies that describe how one set of component failures can trigger another.
Given these constraints, we create generalization and refinement operators that allow us to relax or add constraints as needed.
Thus, we can simplify a system for easier evaluation by generalizing it or we can iteratively develop one through repeated refinement.
Finally, we link these constraints to MIS reliability models, enabling us to refine or generalize models of a common modeling formalism.

We plan to extend the refinement framework to other modeling formalisms and to define higher-level modeling operations, e.g., model composition, in terms of refinement.
Furthermore, we intend to implement this as a software tool so that it can be easily applied to large-scale systems.

\bibliographystyle{ieeetr}
\bibliography{refs}

\end{document}